\documentstyle[amssymb,a4,12pt,epsfig]{article}

\newcommand{\beq}{\begin{equation}}
\newcommand{\eeq}{\end{equation}}
\newcommand{\bea}{\begin{eqnarray}}
\newcommand{\eea}{\end{eqnarray}}

\begin{document}

\title{{\bf Self-dual vortices in a Maxwell Chern-Simons model with non-minimal
coupling}}
\author{{H.R. Christiansen, M.S. Cunha, J.A. Helay\"{e}l-Neto} \\
{L.R.U. Manssur, A.L.M.A. Nogueira}\\
\\
{\normalsize {\it Centro Brasileiro de Pesquisas F\'{\i }sicas, CBPF - DCP
\thanks{Electronic-addresses: hugo,marcony,helayel,leon,nogue@cat.cbpf.br}}}\\
{\normalsize {\it Rua Dr. Xavier Sigaud 150, 22290-180
Rio de Janeiro, Brazil}}}

\date{}
\maketitle

\begin{center}
{\tt To be published in the}\ \ {\it Int.~J.~of~Mod.~Phys.~A\, (1999).}
\end{center}

\begin{abstract}
\noindent
We find self-dual vortex solutions in a Maxwell-Chern-Simons model with
anomalous magnetic moment. From a recently developed $N=2$ supersym\-metric
extension, we obtain the proper Bogomol'nyi equations together with a Higgs
potential allowing both topological and non-topological phases in the theory.
\end{abstract}

\newpage

\section{ Introduction}

A few years ago, it was proposed a Maxwell-Chern-Simons (MCS) gauge theory
with an additional magnetic moment interaction\cite{anomalo} for which
Bogomol'nyi-type self-dual equations can be derived and vortex-like
configurations appear whenever suitable relationships among the parameters
of the model are obeyed \cite{lee}. An important issue that comes about is
the claim of a relation between the property of self-duality and the
$N=2$ supersymmetric extension of the model, accomplished by means of a
relationship between the central charge of the extended model and the
existence of topological quantum numbers \cite{spector}. Although a
fundamental reason for this connection has not been given so far in the
literature, in certain cases it appears to be unavoidable to construct the
$N=2$ supersymmetric extension of a given bosonic model in order to obtain the
proper Higgs potential and self-dual conditions compatible with the
Euler-Lagrange equations.

In this regard, we have succeeded in deriving an $N=2$ Maxwell-Chern-\-Simons
model with anomalous magnetic moment \cite{nos}. Our strategy consisted in
the formulation of an $N=1$ $D=4$ gauge model with a BF-term, free of
constraints on the coupling constants\footnote{
This is to be compared with the procedure of Ref.\cite{navratil} which, in
turn, relies on a special choice of parameters, in order to have an extended
supersymmetry built up directly in $D=3$ dimensions. Similar constraints have
also been needed in order to find an $N=2$ susy extension of the Maxwell
Higgs model \cite{laplata} and of the Chern-Simons Higgs model \cite{llw}.}.
Upon a convenient dimensional reduction of the component-field action from
(1+3) to (1+2) dimensions, we set out an $N=2$ $D=3$ Maxwell-Higgs model with a
Chern-Simons term and magnetic moment interaction with the matter sector.
Adopting this viewpoint, we raised the possibility of freely handling the
parameters of the model and, remarkably, it enabled us to obtain topological
self-dual solutions, even in the critical regime mentioned above. This is to
be compared with previous attempts where just a $\phi ^2$ Higgs potential
has been considered so as to find self-dual solutions \cite{torres}.

In the present paper, we derive the proper self-dual equations and the Higgs
potential needed to allow topological as well as non-topological vortices in
a non-minimally coupled MCS model; this is our main result. We perform a
gauge-independent calculation which permits a suitable handling of the
energy functional, leading to self-dual solutions to the equations of motion
in both the symmetric and asymmetric phases of the model (Sections 2 and 3).
In Section 4, we discuss the properties of system for the critical value of
the magnetic coupling. The analysis of the self-dual solutions and a wide
variety of soliton configurations are presented in Section 5. Finally, in
Section 6, we draw our Conclusions.

%%%%%%%%%%%%%%%%%%%%%%%%%%%%%%%%%%%%%%%%%%%%%%%%%%%%%%%%%%%%%%%%%%%%%%%%%%%

\section{The Lagrangian}

%%%%%%%%%%%%%%%%%%%%%%%%%%%%%%%%%%%%%%%%%%%%%%%%%%%%%%%%%%%%%%%%%%%%%%%%%%%

In Ref.\cite{nos} we have put forward the $N=2$ susy Lagrangian including the
bosonic model we are going to analyse here. In component-field form, it
exhibits the proper non-minimally coupled MCS extension needed for our main
purpose of finding a topological phase in the bosonic theory. For the sake
of a better understanding let us quote below the full expression of the
$N=2$ susy Lagrangian in terms of components

\begin{eqnarray}
{\cal S}_{MCS}^{N=2} &=&\int d^3x\left\{ -\frac 14F_{\mu \nu }F^{\mu \nu
}+\frac \kappa 2\varepsilon ^{\mu \nu \alpha }A_\mu \partial _\nu A_\alpha
+2\Delta ^2+\frac 12\partial _\mu M\partial ^\mu M-2\kappa M\Delta \right.
\nonumber \\
&&+\frac 12\overline{\Lambda }_{-}(i\partial \!\!\!/\,+\kappa )\Lambda _{-}+%
{\rm e}^{2gM}\Bigg[ \nabla _\mu \varphi \nabla ^\mu \varphi
^{*}-(eM+2g\Delta )^2\,\varphi \varphi ^{*}  \nonumber \\
&&+\frac 14(eM+2g\Delta )\,\overline{X}_{+}X_{+}+\frac i8(\overline{X}%
_{-}\nabla \!\!\!\!/\,X_{-}+\overline{X}_{+}\nabla \!\!\!\!/\,X_{+})
\nonumber \\
&&-\frac{ig}2\left( \overline{\Lambda }_{+}\nabla \!\!\!\!/\,\varphi
X_{-}-h.c.\right) +\frac g2(eM+2g\Delta )\left( \overline{\Lambda }%
_{+}X_{-}\varphi +h.c.\right)  \nonumber  \label{eq41} \\
&&+\frac{ig^2}2\partial _\mu M\left( \overline{X}_{-}\gamma ^\mu \Lambda
_{+}\varphi ^{*}-\overline{\Lambda }_{+}\gamma ^\mu X_{-}\varphi \right) -i%
\frac{g^2}2\varphi ^{*}\varphi \left( \overline{\Lambda }_{+}\partial
\!\!\!/\,\Lambda _{+}+\overline{\Lambda }_{-}\partial \!\!\!/\,\Lambda
_{-}\right)  \nonumber \\
&&-\frac{g^2}e\left( \frac 12(\overline{\Lambda }_{+}\gamma ^\mu {\cal J}%
_\mu \Lambda _{+}-\overline{\Lambda }_{-}\gamma ^\mu {\cal J}_\mu \Lambda
_{-})-\overline{\Lambda }_{+}\Lambda _{+}e(eM+2g\Delta )\varphi \varphi
^{*}\right)  \nonumber \\
&&-\varphi \varphi ^{*}\left( 2e\Delta -2ge\overline{\Lambda }_{+}\Lambda
_{+}+g^2\partial _\mu M\partial ^\mu M\right)  \nonumber \\
&&+\left. \left. \frac e2(\varphi \overline{\Lambda }_{+}X_{-}+\varphi ^{*}%
\overline{X}_{-}\Lambda _{+})+\left| S-\frac g2\overline{X}_{-}\Lambda _{-}-%
\frac{g^2}2\varphi \overline{\Lambda }_{+}\Lambda _{-}\right| ^2\Bigg]\right.
\right\}  \label{full3d}
\end{eqnarray}
%%%%%
where %%%%%
\begin{equation}
\nabla _\mu \varphi =\left( \partial _\mu -ieA_\mu -igF_\mu \right) \varphi .
\end{equation}
%% ($\nabla\!\!\!\!/\,_{-}$ means that $i$ is to be changed to $-i$).
The origin of all the fields appearing in equation eq.(\ref{full3d}) has
been carefully justified in \cite{nos}; we refer the reader to this
reference for the details. Here, we are basically concerned with the bosonic
sector of the theory, so we will focus our attention to a particular piece
as we discuss in what follows.

Let us consider the purely bosonic part of the susy Lagrangian of eq.(\ref
{full3d})
\begin{eqnarray}
{\cal L} &=&-\frac 14F_{\mu \nu }^2+\frac 12G\partial _\mu M\partial ^\mu M+%
{\rm e}^{2gM}\nabla _\mu \varphi \nabla ^\mu \varphi ^{*}+\frac \kappa
2A_\mu F^\mu  \nonumber \\
&&+\left\{ 2\Delta ^2-2\kappa M\Delta +\eta \Delta -{\rm e}^{2gM}|\varphi
|^2\left[ (eM+2g\Delta )^2\,+2e\Delta \right] \right\} ,  \label{lag}
\end{eqnarray}
where %%%
\begin{equation}
G(\varphi )=1-2g^2{\rm e}^{2gM}|\varphi |^2  \label{G}
\end{equation}
and  $\eta \Delta $ corresponds to the Fayet-Iliopoulos term
included in the susy Lagrangian in order to allow spontaneous breaking of
gauge invariance \cite{sohnius}.

The equation of motion for the auxiliary $\Delta $--field gives
\begin{equation}
\Delta =\frac e{4G}\left( 2{\rm e}^{2gM}|\varphi |^2-v^2+\frac{2\kappa }eM+4g%
{\rm e}^{2gM}|\varphi |^2M\right)  \label{delta}
\end{equation}
where for convenience we have written $\eta =-ev^2.$ Substitution of the
above in eq.(\ref{lag}), gives the following Higgs-type potential
\begin{equation}
U=\frac{e^2}{8G}\left( 2{\rm e}^{2gM}|\varphi |^2-v^2+\frac{2\kappa }eM+4g%
{\rm e}^{2gM}|\varphi |^2M\right) ^2+e^2M^2{\rm e}^{2gM}|\varphi |^2
\label{higgs1}
\end{equation}
which depends on two fields: a real ($M$) and a complex ($\varphi $) scalar.
Upon elimination of the auxiliary field $\Delta $, we will work with the
following Lagrangian

\begin{equation}
{\cal L}=-\frac 14F_{\mu \nu }^2+\frac 12G\partial _\mu M\partial ^\mu M+%
{\rm e}^{2gM}\nabla _\mu \varphi \nabla ^\mu \varphi ^{*}+\frac \kappa
2A_\mu F^\mu -U  \label{lag2}
\end{equation}
which shall play a central role in the present discussion. Let us first
define the currents
\begin{eqnarray}
H_\mu &=&-\frac{ie}2\left( \varphi ^{*}D_\mu \varphi -\varphi D_\mu \varphi
^{*}\right)  \nonumber \\
{\cal J}_\mu &=&-\frac{ie}2\left( \phi ^{*}\nabla _\mu \phi -\phi \nabla
_\mu \phi ^{*}\right)  \label{jota}
\end{eqnarray}
where $\phi $ is a complex scalar parametrized in terms of $M$ and $\varphi $
as given below
\begin{equation}
\phi =\sqrt{2}{\rm e}^{gM}\varphi .  \label{master}
\end{equation}

As we shall discuss in the next section, the scalar $\phi $ will be
identified as the physical field in terms of which the vortices will be
specified. Now, the equation of motion for the gauge field can be written as
\begin{equation}
\partial _\mu F^{\mu \rho }+\kappa F^\rho ={\cal J}^{\;\rho }+\frac
ge\varepsilon ^{\mu \nu \rho }\partial _\mu {\cal J}_\nu  \label{motion}
\end{equation}
where the time component determines the modified ``Gauss Law''
\begin{equation}
\partial _iE_i+\kappa B+\frac ge\varepsilon _{ij}\partial _i{\cal J}_j+{\cal %
J}_0=0.  \label{gauss}
\end{equation}

The gauge invariant modes are now short-range due to the mass term resulting
from eq.(\ref{gauss}). Hence, the first term has a vanishing space integral.
On the other hand, the third term results in a line integral taken at
infinity which also vanishes for finite energy configurations. Therefore, it
can be seen from the remaining piece that the charge of the vortex solutions
is related to non-zero magnetic fluxes by
\begin{equation}
Q=\kappa \Phi _B,  \label{Q}
\end{equation}
where $\Phi _B\equiv -\int d^2xB.$

%%%%%%%%%%%%%%%%%%%%%%%%%%%%%%%%%%%%%%%%%%%%%%%%%%%%%%%%%%%

\section{The self-dual equations of motion}

%%%%%%%%%%%%%%%%%%%%%%%%%%%%%%%%%%%%%%%%%%%%%%%%%%%%%%%%%%%

The energy functional is given by
\begin{eqnarray}
{\cal E} &=&\int d^2x\left\{ \frac 12G\left( B^2+E^2\right) +\frac
12G\partial _0M\partial _0M+\frac 12G\partial _iM\partial _iM\right.
\nonumber \\
&&\left. \;\;\;\;+{\rm e}^{2gM}D_0\varphi D_0^{*}\varphi +{\rm e}%
^{2gM}D_i\varphi D_i\varphi ^{*}+U\right\}
\end{eqnarray}
which can be reorganized as %%%
\begin{eqnarray}
{\cal E} &=&\int d^2x\left\{ \frac 12G\left( B\mp \frac e{2G}\left( 2{\rm e}%
^{2gM}|\varphi |^2-v^2+\frac{2\kappa }eM+4g{\rm e}^{2gM}|\varphi |^2M\right)
\right) ^2\right.  \nonumber \\
&&\,\pm eB\left( {\rm e}^{2gM}|\varphi |^2-\frac 12v^2+\frac \kappa eM+g{\rm %
e}^{2gM}|\varphi |^2M\right) +\frac 12G\left( E_i\pm \partial _iM\right) ^2
\nonumber \\
&&\,\mp GE_i\partial _iM+{\rm e}^{2gM}\left.\Bigg( \left| D_0\varphi \mp
ieM\varphi \right| ^2\pm 2MH_0+\left| \left( D_1\pm iD_2\right) \varphi
\right| ^2\right.  \nonumber \\
&&\left. \left. \,\pm \frac 1e\varepsilon _{ij}\partial _iH_j\mp eB|\varphi
|^2\right.\Bigg) \right\}
\end{eqnarray}

Notice that the non-minimal term from the $\nabla _\mu $--derivative, though
not explicitly written in the above equation, has an effect which is
implicit through $G$. The terms linear in $F_\mu $ are not present in the
energy because they are metric-independent.

Now, the search of the Bogomol'nyi bound for the energy yields the proper
self-dual equations in a natural way
\begin{eqnarray}
&&B\mp \frac e{2G}\left( 2{\rm e}^{2gM}|\varphi |^2-v^2+\frac{2\kappa }eM+4g%
{\rm e}^{2gM}|\varphi |^2M\right) =0  \nonumber \\
&&E_i\pm \partial _iM=0  \nonumber \\
&&D_0\varphi \mp ieM\varphi =0  \nonumber \\
&&\left( D_1\pm iD_2\right) \varphi =0.  \label{dual}
\end{eqnarray}
Using the following identities
\begin{eqnarray*}
\frac 1e{\rm e}^{2gM}\varepsilon _{ij}\partial _iH_j &=&\frac
1{2e}\varepsilon _{ij}\partial _i{\cal J}_j+\frac 1{2e}\partial _iE_i-\frac
ge\varepsilon _{ij}\left( \partial _iM\right) {\cal J}_j-2g^2{\rm e}%
^{2gM}|\varphi |^2E_i\partial _iM \\
2{\rm e}^{2gM}H_0 &=&{\cal J}_0-eg{\rm e}^{2gM}|\varphi |^2,
\end{eqnarray*}
integrating by parts and dropping surface terms, one finally gets
\begin{eqnarray}
{\cal E} &=&\frac{ev^2}2|\Phi _B|+\int d^2x\left\{ \pm M\left( {\cal J}%
_0+\frac ge\varepsilon _{ij}\partial _i{\cal J}_j+\kappa B+\partial
_iE_i\right) +\frac 12G\partial _0M\partial _0M\right.  \nonumber \\
&&\left. \pm \frac 1{2e}\varepsilon _{ij}\partial _i{\cal J}_j\pm \frac
1{2e}\partial _iE_i\right\} .  \nonumber
\end{eqnarray}
The last two terms vanish whenever integrated over the whole space; so,
using the Gauss law and considering static configurations, the lower bound
to the energy is clearly attained.

To close this section, let us express the self-dual equations and the Higgs
potential in terms of the $\phi $--field; in so doing, we shall get
expressions that are more useful for our future purposes. Now,
eqs. (\ref{dual}) and (\ref{higgs1}) read
\begin{eqnarray}
B\mp \frac e{2G}\left( |\phi |^2-v^2+\frac{2\kappa }eM+2g|\phi |^2M\right)
&=&0  \nonumber \\
A_0\pm M &=&0  \nonumber \\
\nabla _1\phi \pm i\nabla _2\phi &=&0,  \label{dual2}
\end{eqnarray}
and
\begin{equation}
U=\frac{e^2}{8G}\left( |\phi |^2-v^2+\frac{2\kappa }eM+2g|\phi |^2M\right)
^2+\frac{e^2}2M^2|\phi |^2  \label{higgs2}
\end{equation}
which, for $g=0$, gives the Higgs potential of the minimal MCS model as
given by the supersymmetric Lagrangian found in ref.\cite{lee}, as expected.
Notice that the system has two degenerate minima, a symmetric phase for
which $\left| \phi \right| =v,$ $M=0$ and an asymmetric phase where $\phi
=0, $ $M=ev^2/2\kappa .$

%%%%%%%%%%%%%%%%%%%%%%%%%%%%%%%%%%%%%%%%%%%%%%%%%%%%%%%%%%%

\section{The critical magnetic coupling}

%%%%%%%%%%%%%%%%%%%%%%%%%%%%%%%%%%%%%%%%%%%%%%%%%%%%%%%%%%%

Let us now analyze a very special value of the magnetic coupling, namely,
\begin{equation}
g_c=-e/\kappa ,
\end{equation}
for which the equations of motion (\ref{motion}) reduce down to first order,
looking similar to the pure CS model's. This choice yields fractional
statistics describing anyons \cite{anomalo}. Remarkably enough, this is the
value that has to be fixed in order to obtain an $N=2$ MCS non-minimal
theory, when working from the outset in $D=3$ \cite{navratil}.
It is important to notice that by performing the susy extension without
dimensional reduction,
only a symmetric, $\phi ^2$, Higgs potential has been found,
yielding, consequently, just non-topological solutions \cite{torres}.

Hence, for $g=g_c$ one has
\begin{equation}
{\cal J}_\mu =\kappa F_\mu
\end{equation}
whose time-component reads
\begin{equation}
\kappa \left( 1-\frac{e^2}{\kappa ^2}|\phi |^2\right) B=e^2A_0|\phi |^2.
\end{equation}

We will now show that, in our model, we can make such a special choice,
$g=g_c$, without constraining the potential to a symmetric phase. We shall
also find topological vortices, in contrast to previous attempts (we drop
the subscript $_c$ in what follows).

Using $A_0=\mp M$ [see eq.(\ref{dual2})] and defining $\gamma $ by means of $%
\kappa =\gamma ev$ $(\gamma \geq 1$ to have positive definite energy
configurations) we get

\begin{equation}
B=\mp \gamma evM\frac{|\phi |^2}{\gamma ^2v^2-|\phi |^2}.  \label{b1}
\end{equation}
On the other hand, self-duality relations provide also
\begin{equation}
B=\pm \frac{ev^2}2\frac{|\phi |^2-v^2}{\gamma ^2v^2-|\phi |^2}\gamma ^2\pm
\gamma evM.  \label{b2}
\end{equation}
Thus, for $v$ defined as the maximum value of $|\phi |$, eqs. (\ref{b1}) and
(\ref{b2}) give

\begin{eqnarray}
M &=&-\frac{\left( |\phi |^2-v^2\right) }{2\gamma v}  \nonumber \\
B &=&\pm \frac e2\frac{|\phi |^2\left( |\phi |^2-v^2\right) }{\left( \gamma
^2v^2-|\phi |^2\right) }.  \label{bm}
\end{eqnarray}

Notice that the $M$ field has decoupled from the other components, i.e., it
can be written in terms of just the Higgs field $\phi $. Another interesting
feature of the bosonic model just found is that it still presents together
both topological and non-topological phases
\begin{equation}
U=\frac{e^4}8\frac{|\phi |^2(|\phi |^2-v^2)^2}{{\kappa ^2}-e^2|\phi |^2}
\label{higgscrit}
\end{equation}
Note that for $\kappa \rightarrow \infty $ $(\gamma >>1),$ it behaves like
the Higgs potential typical of a pure CS model \cite{llw}, as expected.

%%%%%%%%%%%%%%%%%%%%%%%%%%%%%%%%%%%%%%%%%%%%%%%%%%%%%%%%%%%%%%%%%%%%%%%%%

\section{Analysis of the self-dual solutions}

%%%%%%%%%%%%%%%%%%%%%%%%%%%%%%%%%%%%%%%%%%%%%%%%%%%%%%%%%%%%%%%%%%%%%%%%%

Assuming maximal (rotational) symmetry, we take the following ansatz to find
self-dual vortices
\begin{eqnarray}
\phi (r,\theta ) &=&vR(r){\rm e}^{in\theta }  \label{ansatz1} \\
{\bf A}(r) &=&\frac{\widehat{\theta }}{er}\left[ a(r)-n\right]
\label{ansatz2}
\end{eqnarray}
where $R$ and $a$ are real functions of $r$, and $n$ an integer indicating
the topological charge of the vortex. Then, the magnetic field reads
\begin{equation}
B=\frac 1{er}a^{\prime }
\end{equation}
and the flux is
\begin{equation}
\Phi _B=\frac{2\pi }e\left[ a(0)-a(\infty )\right]
\end{equation}

Now, since in polar coordinates one has
\begin{equation}
\partial _1\pm i\partial _2={\rm e}^{\pm i\theta }\left( \partial _r\pm
\frac ir\partial _\theta \right) ,
\end{equation}
eq. (\ref{dual2}) can be written as %%%
\begin{eqnarray}
\left( 1-\frac{R^2}{\gamma ^2}\right) \frac{dR}{dr}\mp \frac arR &=&0
\label{dual_aR1} \\
\frac 1r\frac{da}{dr}\mp \frac{R^2\left( R^2-1\right) }{\left( \gamma
^2-R^2\right) } &=&0  \label{dual_aR2}
\end{eqnarray}
where we have used (\ref{bm}) and redefined $r\rightarrow \frac{\sqrt{2}}{ev}%
r.$

The natural boundary conditions at infinity result from the requirement of
finite energy, while a non-singular behavior determines the values at the
origin.

In the {\it topological} phase, $R(\infty )=1$ and $a(\infty )=0$ for
nontrivial vorticity $n$. Then, for large $r$ the asymptotic form of the
topological vortices is given by
\begin{eqnarray}
R(r) &\simeq &1-\frac{\sqrt{2}}2d\ \gamma K_0\left( \frac{\sqrt{2}}{\gamma
^2-1}\gamma r\right)  \label{topolarge1} \\
a(r) &\simeq &d\ rK_1\left( \frac{\sqrt{2}}{\gamma ^2-1}\gamma r\right)
\label{topolarge2}
\end{eqnarray}
where $d$ is a constant whose value is determined by the form of the
solutions at the origin. Also at the origin, one must expect non-singular
fields, implying $R(0)=0$ and $a(0)=n$. Hence, the magnetic flux is
quantized as follows
\begin{equation}
\Phi _B=\frac{2\pi }e\left[ a(0)-a(\infty )\right] =\frac{2\pi }en.
\end{equation}

Now, we can combine eqs. (\ref{dual_aR1},\ref{dual_aR2}) to produce a second
order equation
\begin{equation}
\frac 1r\frac d{dr}(r\frac{dR}{dr})=\frac{\left( R^{\prime }\right) ^2}R%
\frac{1+R^2/\gamma ^2}{1-R^2/\gamma ^2}-\frac{1-R^2}{\gamma ^2\left(
1-R^2/\gamma ^2\right) ^2}R^3  \label{sec_ord}
\end{equation}
so that the behavior of the solutions for small values of $r$, where $(1\pm
R^2)\simeq 1$, can be approximated by a Liouville-type function
\begin{equation}
R(r)\simeq \frac{2N\gamma }r\left[ \left( \frac r{r_0}\right) ^N+\left(
\frac{r_0}r\right) ^N\right] ^{-1}  \label{liousol}
\end{equation}
where $N$ and $r_0$ are arbitrary constants. Upon substitution of the above
expression in eq.(\ref{dual_aR1}), we find
\begin{equation}
a(r)\simeq -1+N\frac{1-\left( \frac r{r_0}\right) ^N}{1+\left( \frac
r{r_0}\right) ^N},  \label{asolut}
\end{equation}
so that using $a(0)=n$ we obtain $N=n+1$. It implies that, near the origin,
the form of the vortex is power-like %%
\begin{eqnarray}
R(r) &\simeq &\gamma c_nr^n,  \nonumber \\
\ \ \ \ \ a(r) &\simeq &n-c_nr^{n+1}  \label{orig}
\end{eqnarray}
where $c_n=2(n+1)/r_0^{n+1}$. This last relation is obtained by expanding
eq.(\ref{liousol}) around $r=0$; however, the precise numerical values of
the $c_n$ constants are determined by the shape of the fields at infinity,
rather than by their behavior at the origin. Indeed, we have numerically
solved the self-dual equations of motion by means of an iterative procedure,
giving a tentative value for $c_n$ which is corrected each time by imposing
that both limits, $R\rightarrow 1$ and $a\rightarrow 0$, hold together at
infinity. For illustration, we quote some of the results in Fig. 1 and Fig.
2, for the cases $n=1,2,3.$ Notice the ring-type structure of the
topological vortices (see. Fig. 2 and Fig. 3). This profile is analogous to
the pure CS magnetic field shape \cite{jw}. The $n<0$ configurations are
related to the $n>0$ ones by the transformation $a\rightarrow -a$ and $%
R\rightarrow R$. \strut

\begin{center}
Table 1. Several values of $c_n$ for $n=1,2,3$ and $\gamma =1.5,2.0,4.0$\
[see eq.(\ref{orig})]. \vskip 0.2cm
\begin{tabular}{|c|c|c|c|}
\hline
$\gamma \diagdown n$ & $1$ & $2$ & $3$ \\ \hline
$1.5$ & $1.7948\times 10^{-1}$ & $2.0538\times 10^{-2}$ & $1.4206\times
10^{-3}$ \\ \hline
$2.0$ & $1.0664\times 10^{-1}$ & $9.1194\times 10^{-3}$ & $4.7256\times
10^{-4}$ \\ \hline
$4.0$ & $2.8118\times 10^{-2}$ & $1.1979\times 10^{-3}$ & $3.1004\times
10^{-5}$ \\ \hline
\end{tabular}
\end{center}

\vskip 0.3cm

Now, let us analyze the {\it non-topological} sector. In this case $v$ is no
longer a relevant parameter, but we can use the same ansatz as in eqs.(\ref
{ansatz1},\ref{ansatz2}) with $v=\kappa /e$ ($\gamma =1$). Now, the system
of differential equations gets simplified by
\begin{eqnarray}
\frac 1r\frac{da}{dr} &=&\mp R^2  \nonumber \\
\frac{dR}{dr} &=&\pm \frac{aR}{r\left( 1-R^2\right) },  \label{dualnontop}
\end{eqnarray}
but it still admits soliton solutions. These are analogous to those found
with the $\phi ^2$ potential considered in \cite{torres}, although in our
case the symmetric phase of the potential arises from $U=\frac{e^2}8|\phi
|^2\left( \frac{\kappa ^2}{e^2}-|\phi |^2\right) $ and the soliton
structures are of course not identical. It should be mentioned that in the
present situation a nontrivial vacuum value, $|\phi |^2=\frac{\kappa ^2}{e^2}
$, is not physically meaningful as the magnetic flux $\Phi _B$ together with
the energy would be divergent. Note also that for large $r$, where $%
(1-R^2)\simeq 1$, the second order equation for the field $R(r)$ becomes
again Liouville's type so that asymptotic solutions look as (\ref{liousol})
and (\ref{asolut}) in contrast to the asymptotic behavior of topological
solutions as given in eqs.(\ref{topolarge1}) and (\ref{topolarge2}). Since
we need $R(\infty )=0$ to have finite energy configurations, from (\ref
{asolut}) we obtain $N^\infty =-(a_n^\infty +1)$, where $a(\infty
)=a_n^\infty $.

At the origin, non-singular soliton configurations must satisfy $nR(0)=0$
and $a(0)=n$. Let us now distinguish between the following two possibilities.

On the one hand, non-vorticity $(n=0$) implies that $R(0)=b_0$ is a
continuous parameter, restricted between $0$ and 1. This restriction is to
guarantee the validity of eq.(\ref{sec_ord}) for all $r$, namely to avoid
singularities.

When $b_0\rightarrow 0,$ we may assume that Liouville's approximation is
valid for all $r,$ and then we can employ (\ref{asolut}) to calculate $%
a_0^\infty $ by using just the value at the origin $a(0)=0.$ It provides $%
(a_0^\infty )_{\min }=-2$ as an analytical result. When $b_0\rightarrow 1,$
we can not use (\ref{asolut}) at the origin any longer, and we have to
perform numerical calculations yielding $(a_0^\infty )^{\max }=-1.83$ (see
Figs. 4,5,6 and Table 2). In Figs. 4-6 it can be seen the flux-tube
structure of the $n=0$ solitons, with the maximum value of the magnetic
field at the origin.

On the other hand, non-trivial vorticity ($n\neq 0$) implies that, at the
origin, $R(r)$ can only be zero. Then we can use (\ref{asolut}) so that
again $N=n+1$. Thus, for $r\simeq 0$ we have
\begin{eqnarray}
&&R(r)\simeq b_nr^n \\
&&a(r)\simeq n-b_nr^{n+1}  \nonumber
\end{eqnarray}
but in contrast to the topological case, the constants are now continuous
parameters bounded as $0<b_n\leq b_n^{\max }$. For $b_n\simeq 0$, $R(r)\ll 1$
and Liouville's approximation is valid everywhere, hence, we are able to
analytically obtain a lower bound for $a(\infty )$ namely, $(a_n^\infty
)_{\min }=-(n+2)$. On the other hand, by numerical investigation we can
obtain the maximum values $b_n^{\max }$ and accordingly $(a_n^\infty )^{\max
}$, as we illustrate in Table 2.

Thus, in the non-topological phase the magnetic flux is not quantized but
instead it is bounded for each vortex number; the width of the band shrinks
as $n$ is increased, varying continuously between
\begin{equation}
2(n-(a_n^\infty )_{\max })\leq \frac e\pi \Phi _B\leq 4(n+1).
\end{equation}

\begin{center}
Table 2. Values of $b_n^{\max }$ and $-(a_n^\infty)_{\max }$ for $%
n=0,1,2,3,4,5$.

\vskip0.2cm

\begin{tabular}{|c|c|c|}
\hline
$n$ & $b_n^{\max }$ & $-(a_n^\infty )^{\max }$ \\ \hline
$0$ & $1.0000$ & $1.8300$ \\ \hline
$1$ & $3.4648\times 10^{-1}$ & $2.5484$ \\ \hline
$2$ & $5.9971\times 10^{-2}$ & $3.5210$ \\ \hline
$3$ & $6.2389\times 10^{-3}$ & $4.5112$ \\ \hline
$4$ & $4.6720\times 10^{-4}$ & $5.5069$ \\ \hline
$5$ & $2.6900\times 10^{-5}$ & $6.5011$ \\ \hline
\end{tabular}
\end{center}

\vskip 0.3cm

As we show in Table 2, the asymptotic values of the gauge field remain
constrained, increasingly close to $-(n+3/2)$ while the magnetic flux
approaches the limit $(4n+3)\frac \pi e$ as $n$ grows.

In Fig. 9, we show the ring structure of the non-topological vortices, with
the maximum of the magnetic field out of the origin as it happens in the
topological phase of the model. Notice that for $b_n^{\max }$ the radius of
the vortices, together with their distance to the origin, are minimal for a
given charge $n$, while the Higgs field presents a cuspidal profile
attaining its maximum value, $R=1^{-}$.

\section{Conclusions}

In this work, we have obtained the self-dual soliton solutions of a
Maxwell--Chern-Simons model with anomalous coupling, in both topological
and non-topological sectors. To
do this, we have focused the bosonic part of a $N=2$ $D=3$ supersymmetric
model --obtained by dimensional reduction from a $N=1$ $D=4$ theory--
which enabled us finding a topological phase in $D=3$. As long
as we know, it is the first time that topological self-dual vortices are
found in a non-minimal MCS system.

We have also analyzed the non-topological
phase in detail for several values of the parameters and magnetic fluxes. It
is worth noting that in contrast to previous reports \cite{torres} the
non-topological phase of our model is not given by a simple $\phi^2$ Higgs
potential but rather as a fourth order function $|\phi |^2(\frac{\kappa ^2}
{e^2}-|\phi |^2)$. Our non-topological solutions are then different from
those presented in \cite{torres} although similar in shape.

In order to compare our results with the literature at hand we have
especially considered the critical anomalous coupling. Remarkably,
we have shown that in this case it is possible to obtain a topological
phase in the non-minimal Maxwell--Chern-Simons model in contrast with
foregoing publications.
We have also found that the corresponding non-topological
phase is not the one analyzed in previous attempts.
A natural extension of this work would be relaxing the anomalous coupling
so as to consider non-critical values of $g$. In principle, such a general
situation could also present topological solitons. Since analytical as well
as computational analyses are more involved in that case, it is still under
investigation and the results shall be soon reported elsewhere.

\section*{Acknowledgments}

The authors would like to thank C. N\'{u}\~{n}ez for useful discussions.
H.R.C. is supported
by FAPERJ--Rio de Janeiro, M.S.C., A.L.N. and J.A.H. by CNPq--Brazil,
and L.U.M. by CAPES--Brazil.
%%%%%%%%%%%%%%%%%%%%%%%%%%% Bibliography %%%%%%%%%%%%%%%%%%%%%%%%%%%%%

%%%%%%%%%%%%%%%%%%%%%%%%%%% Figures-Topological %%%%%%%%%%%%%%%%%%%%%%%%%%%%%

%%%%%%% figure 1
\unitlength=1cm
\begin{figure}[tbp]
\centering
\begin{picture}(10,8)
\epsfig{file=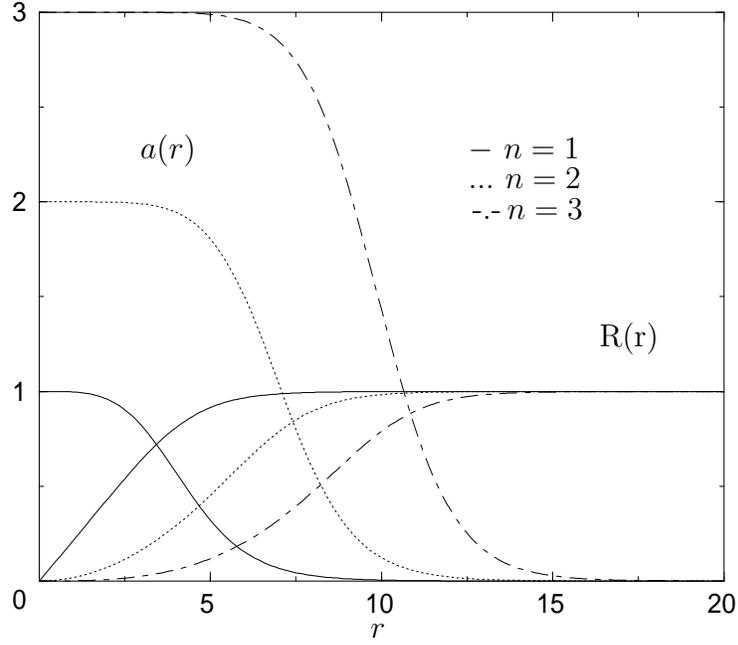,width=10cm,height=08cm}
\put(-3.5,6){$ { \-- \,\, n=1} $}
\put(-3.5,5.6){$ ... \,{ \, n=2} $}
\put(-3.45,5.2){$ $-.-$\,{ n=3 }$}
\put(-8,6){{ $a(r)$}}
\put(-1.9,3.5){{ R(r)}}
\put(-4.8,-.38){$ { r} $}
\end{picture}
\vskip 0.3cm
\caption{The scalar $R(r)$ and the gauge field $a(r)$ in the topological
phase. The values of the $c_n$ constants are fixed by the shape of the
fields at infinity: $c_1=0.1066$, $c_2=9.1190\times 10^{-3}$, $%
c_3=4.726\times 10^{-4}$, for $\gamma =2$.}
\end{figure}
%%%%%%% figure 2
\vskip 1.6cm
\begin{figure}[tbp]
\centering
\begin{picture}(10,8)
\epsfig{file=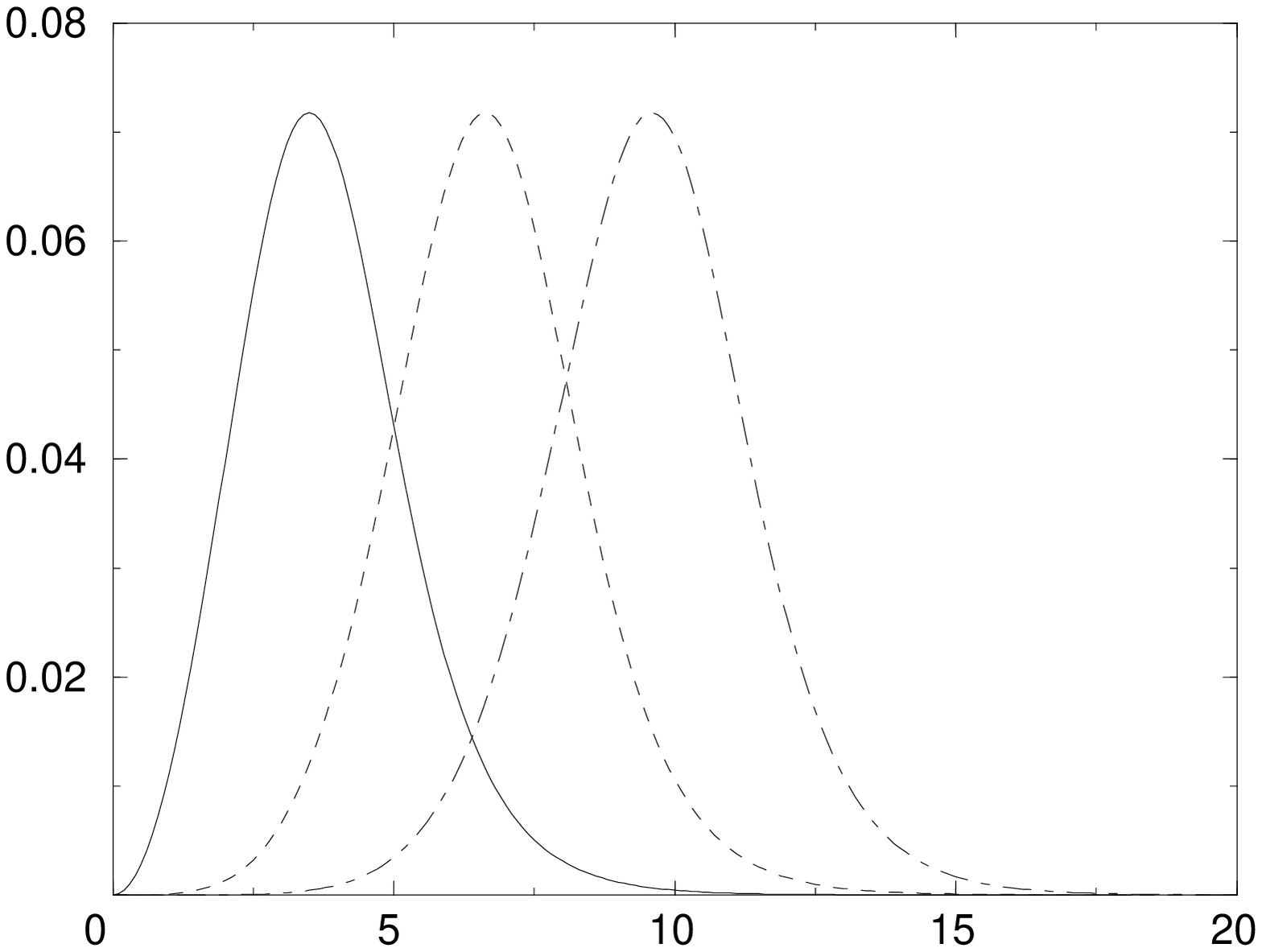,width=10cm,height=08cm}
\put(-4.8,-.38){$ { r} $}
\put(-12.2,4){$
{ -\frac{2}{hv^2} { B(r)} }$}
\put(-3.5,6){$ { \-- \, n=1} $}
\put(-3.5,5.6){$ ... \, { n=2 }$}
\put(-3.45,5.2){$ $-.-${ n=3 }$}
\end{picture}
\vskip 0.3cm
\caption{The magnetic field $B$ as a function of $r$ for $n=$~1, 2, 3 and $%
\gamma =2$ in the topological phase. The vortex structure is ring-type like
pure CS vortices.}
\end{figure}

%%%%%%% figure 3
\begin{figure}[ht]
\centering
\begin{picture}(10,8)
\epsfig{file=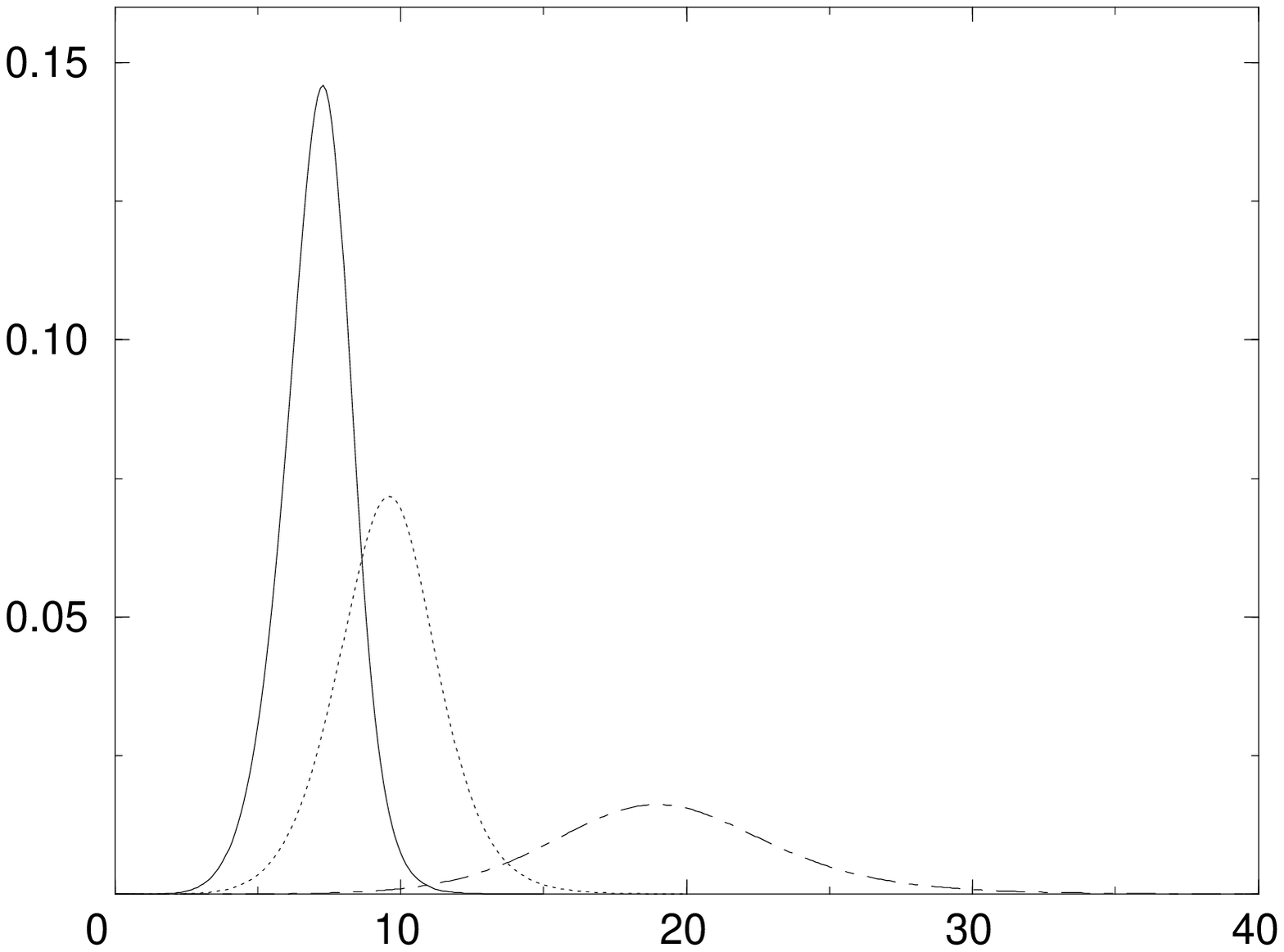,width=10cm,height=08cm}
\put(-3.5,4.8){$ \-- \,{ \gamma=1.5} $}
\put(-3.5,4.3){$ ... \, { \gamma=2.0 }$}
\put(-3.5,3.8){$ $-.-$ \,{\gamma=4.0} $}
\put(-12.2,4){${- \frac{2}{hv^2} { B(r)} }$}
\put(-4.8,-.38){$ { r} $}
\end{picture}
\vskip 0.3cm
\caption{The magnetic field $B$ in the topological phase for $n=1$ and
several values of $\gamma=\frac{\kappa}{hv}$.}
\end{figure}

%%%%%%%%%%%%%%%%%%%%%%%%%%% Endfigures-Topological %%%%%%%%%%%%%%%%%%%%%%

%%%%%%%%%%%%%%%%%%%%%%%%%%% Figures-Nontopological %%%%%%%%%%%%%%%%%%%%%%
%%%%%%%%%%%%%%%%%%%%%%
%%%%%%%%%%%%%%%%%%%%%%  -------- n=0 ---------------
%%%%%%% figure 4
\vskip 1.6cm
\begin{figure}[tbp]
\centering
\begin{picture}(10,8)
\epsfig{file=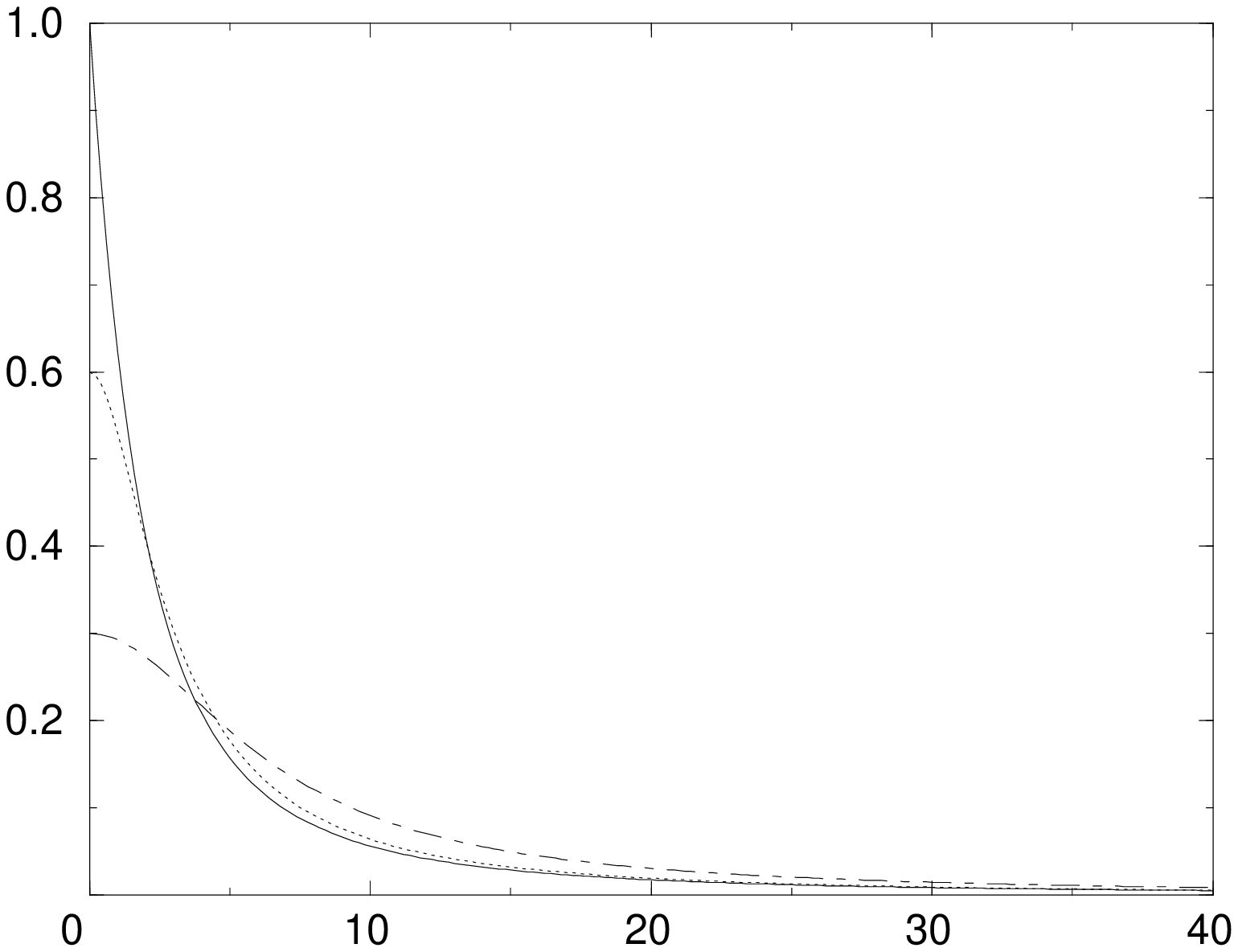,width=10cm,height=08cm}
\put(-3.5,4.8){$ \-- \,{ b_0=0.99} $}
\put(-3.5,4.3){$ ... \, { b_0=0.60 }$}
\put(-3.5,3.8){$ $-.-$ \,{ b_0=0.30} $}
\put(-11.3,4){$ { R(r)} $}
\put(-4.8,-.38){$ { r} $}
\end{picture}
\par
\vskip 0.3cm
\caption{The Higgs field $R(r)$ in the non-topological phase for $n=0$ and
several values of $b_0$.}
\end{figure}

%%%%%%% figure 5
\begin{figure}[tbp]
\centering
\begin{picture}(10,8)
\epsfig{file=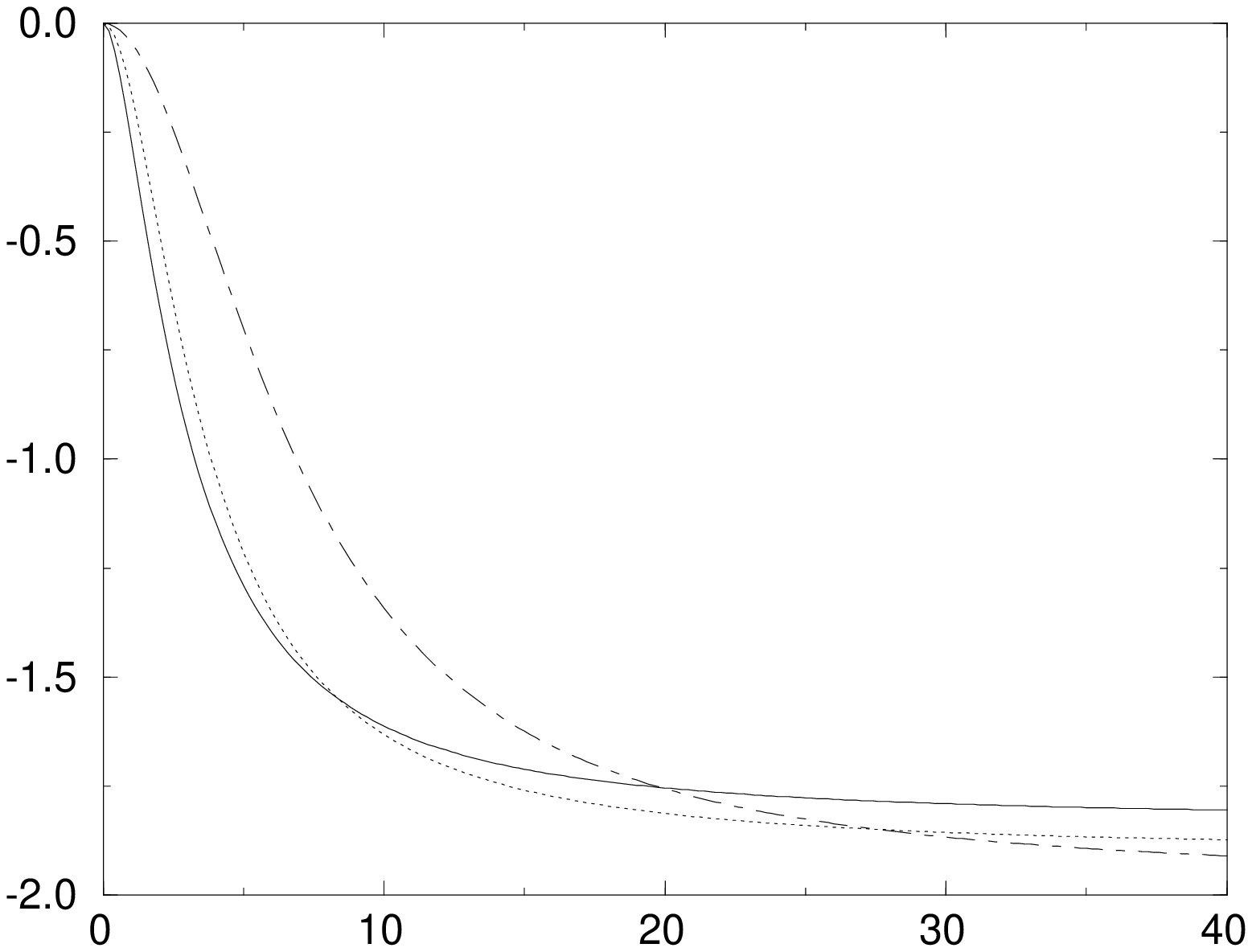,width=10cm,height=08cm}
\put(-3.5,4.8){$ \-- \,{ b_0=0.99} $}
\put(-3.5,4.3){$ ... \, { b_0=0.60 }$}
\put(-3.5,3.8){$ $-.-$ \,{ b_0=0.30} $}
\put(-11.3,4){$ { a(r)} $}
\put(-4.8,-.38){$ { r} $}
\end{picture}
\vskip 0.3cm
\caption{The gauge field $a(r)$ in the non-topological phase for $n=0$ and
several values of $b_0$.}
\end{figure}
%%%%%%%%% figure 6
\vskip 1.5cm
\begin{figure}[tbp]
\centering
\par
\begin{picture}(10,8)
\epsfig{file=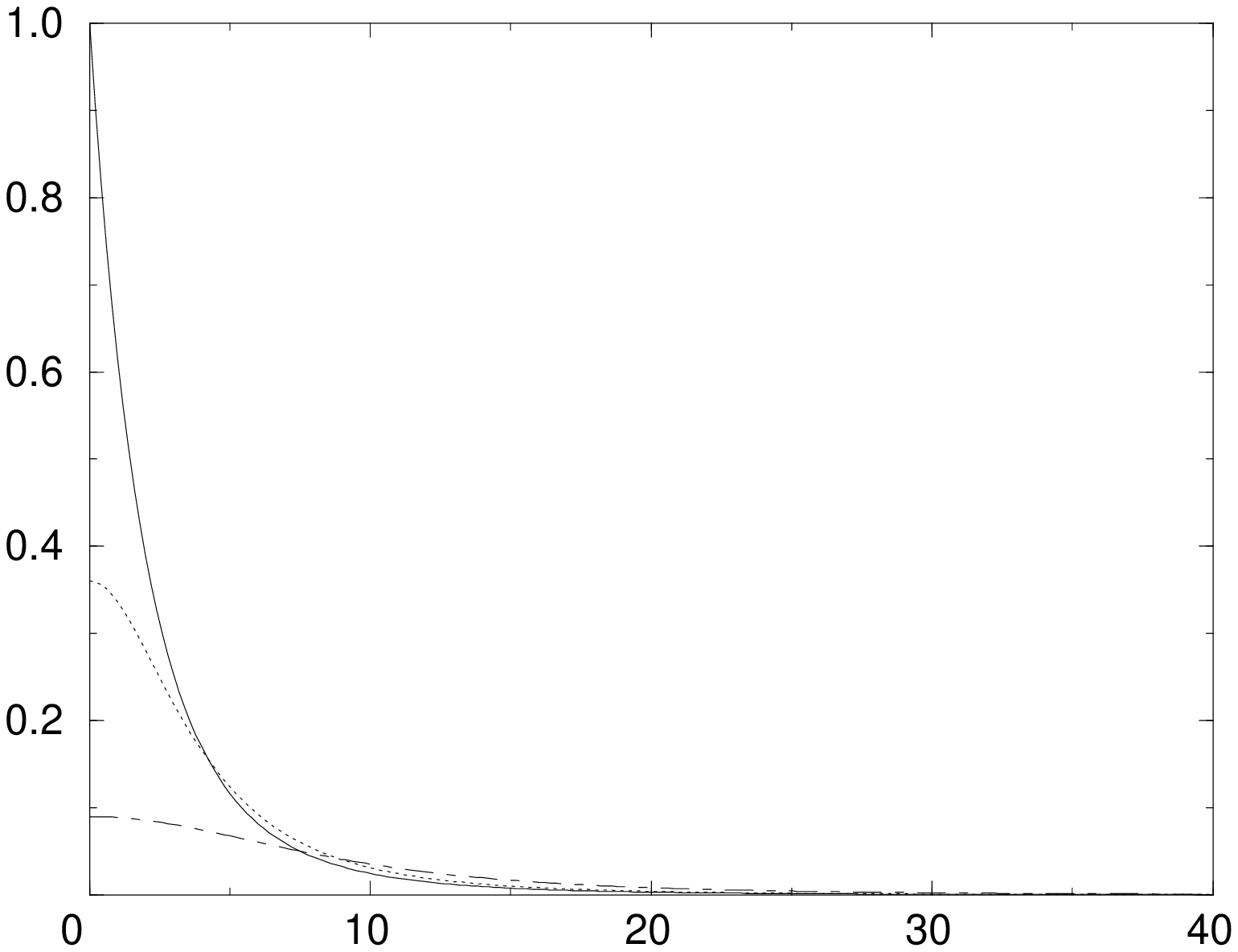,width=10cm,height=08cm}
\put(-3.5,4.8){$ \-- \,{ b_0=0.99} $}
\put(-3.5,4.3){$ ... \, { b_0=0.60 }$}
\put(-3.5,3.8){$ $-.-$ \,{ b_0=0.30} $}
\put(-12.2,4){${- \frac{2}{hv^2} { B(r)} }$}
\put(-4.8,-.38){$ { r} $}
\end{picture}
\vskip 0.3cm
\caption{The magnetic field $B(r)$ in the non-topological phase for $n=0$
and several values of $b_0$. The structure looks as a flux tube.}
\end{figure}

%%%%%%%%%%%%%%%%%%%%%% ---------- n=1,2,3 ----------

%%%%%%% figure 7
\begin{figure}[ht]
\centering
\begin{picture}(10,8)
\epsfig{file=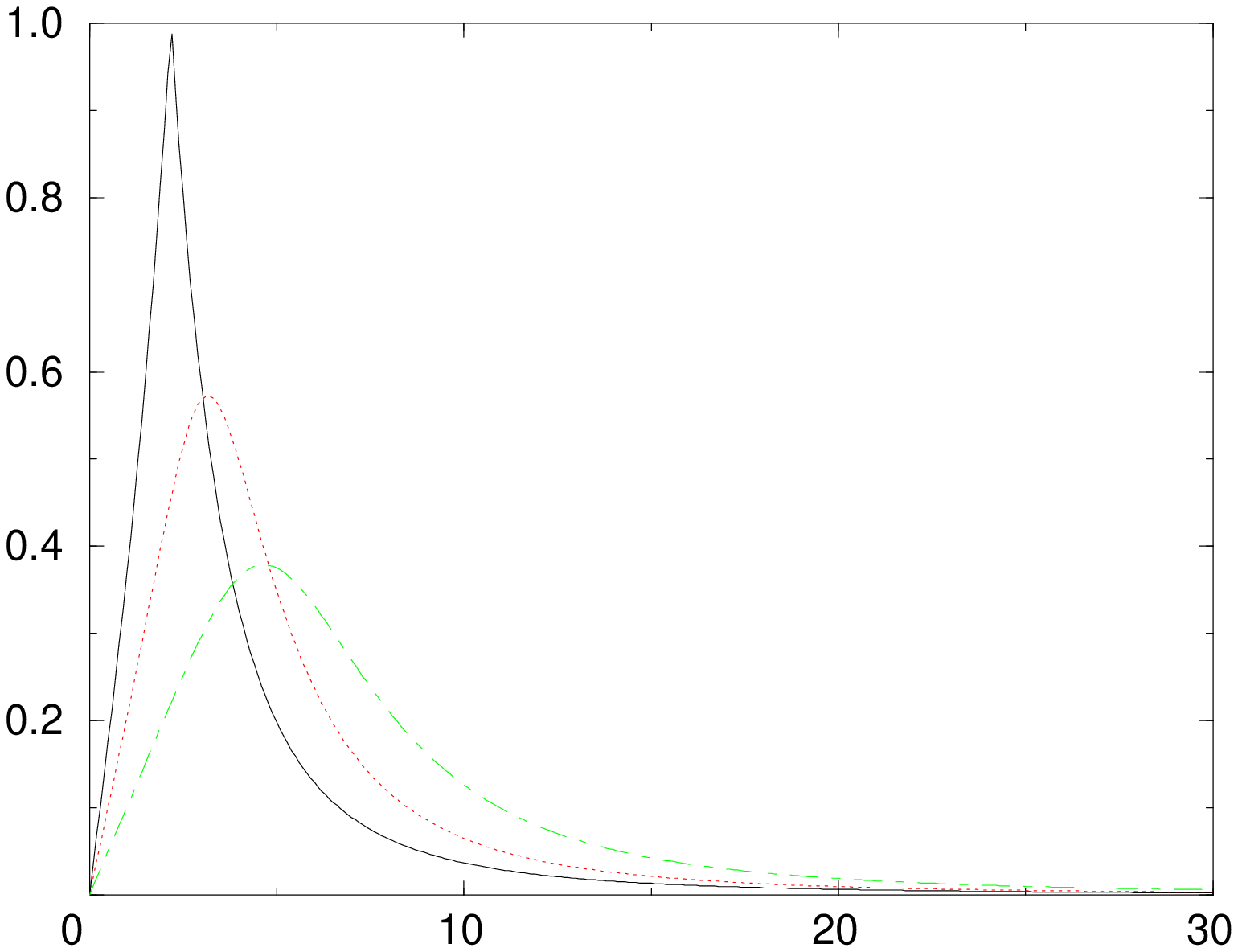,width=10cm,height=08cm}
\put(-5,4.8){$ \-- \,{ b_1 \simeq b_{1max}=0.3465} $}
\put(-5,4.3){$ ... \, { b_1=0.2 }$}
\put(-5,3.8){$ $-.-$ \,{ b_1=0.1} $}
\put(-11.3,4){${ { R(r)} }$}
\put(-4.8,-.38){$ { r} $}
\end{picture}
\vskip 0.3cm
\caption{The Higgs field $R(r)$ in the non-topological phase for $n=1$ and
several values of $b_1$.}
\end{figure}

%%%%%%% figure 8
\vskip 1.6cm
\begin{figure}[tbp]
\centering
\begin{picture}(10,8)
\epsfig{file=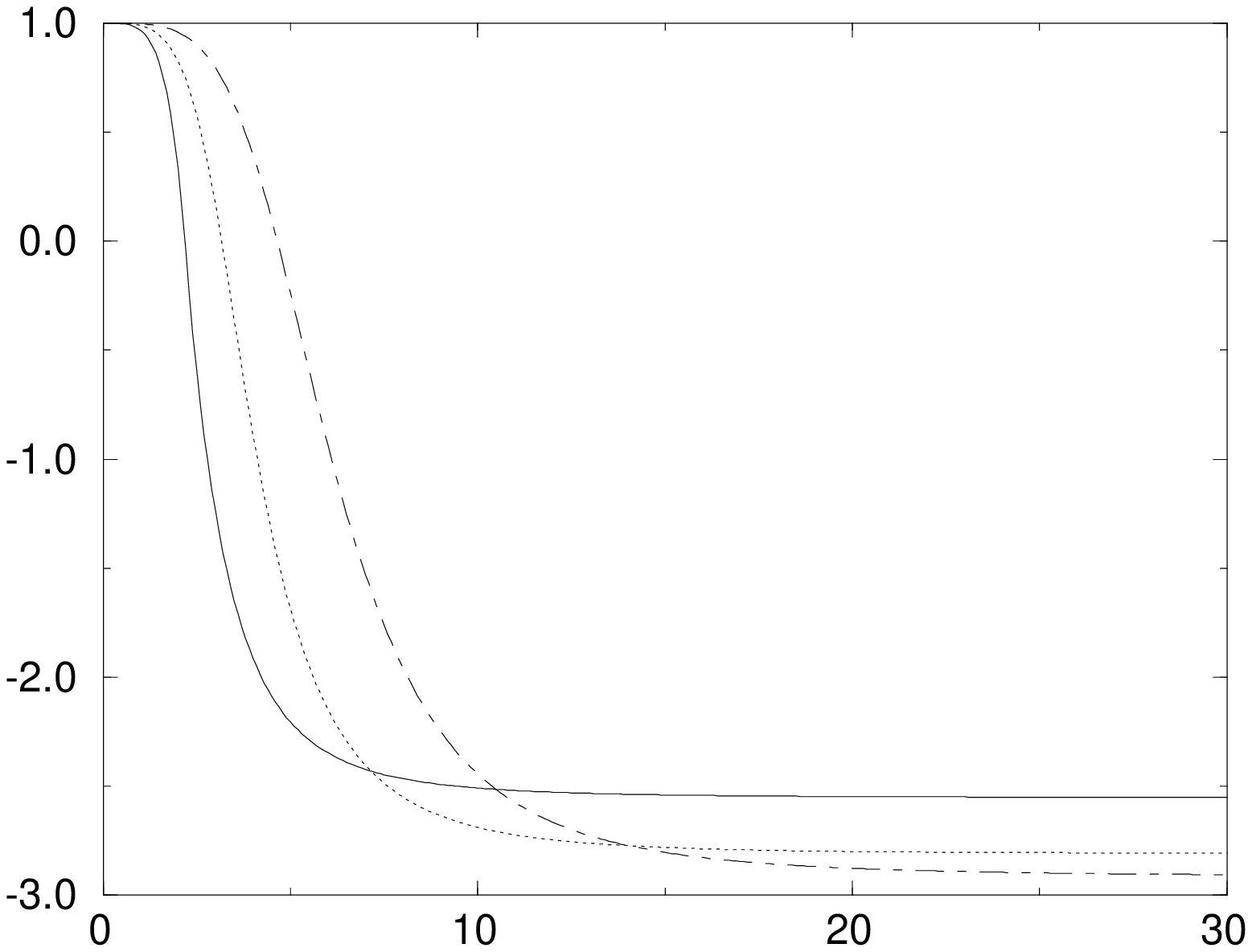,width=10cm,height=08cm}
\put(-5,4.8){$ \-- \,{ b_1 \simeq b_{1max}=0.3465} $}
\put(-5,4.3){$ ... \, { b_1=0.2 }$}
\put(-5,3.8){$ $-.-$ \,{ b_1=0.1} $}
\put(-11.3,4){${ { a(r)} }$}
\put(-4.8,-.38){$ { r} $}
\end{picture}
\vskip 0.3cm
\caption{The gauge field $a(r)$ in the non-topological phase for $n=1$ and
several values of $b_1$.}
\end{figure}

%%%%%%% figure 9

\begin{figure}[ht]
\centering
\begin{picture}(10,8)
\epsfig{file=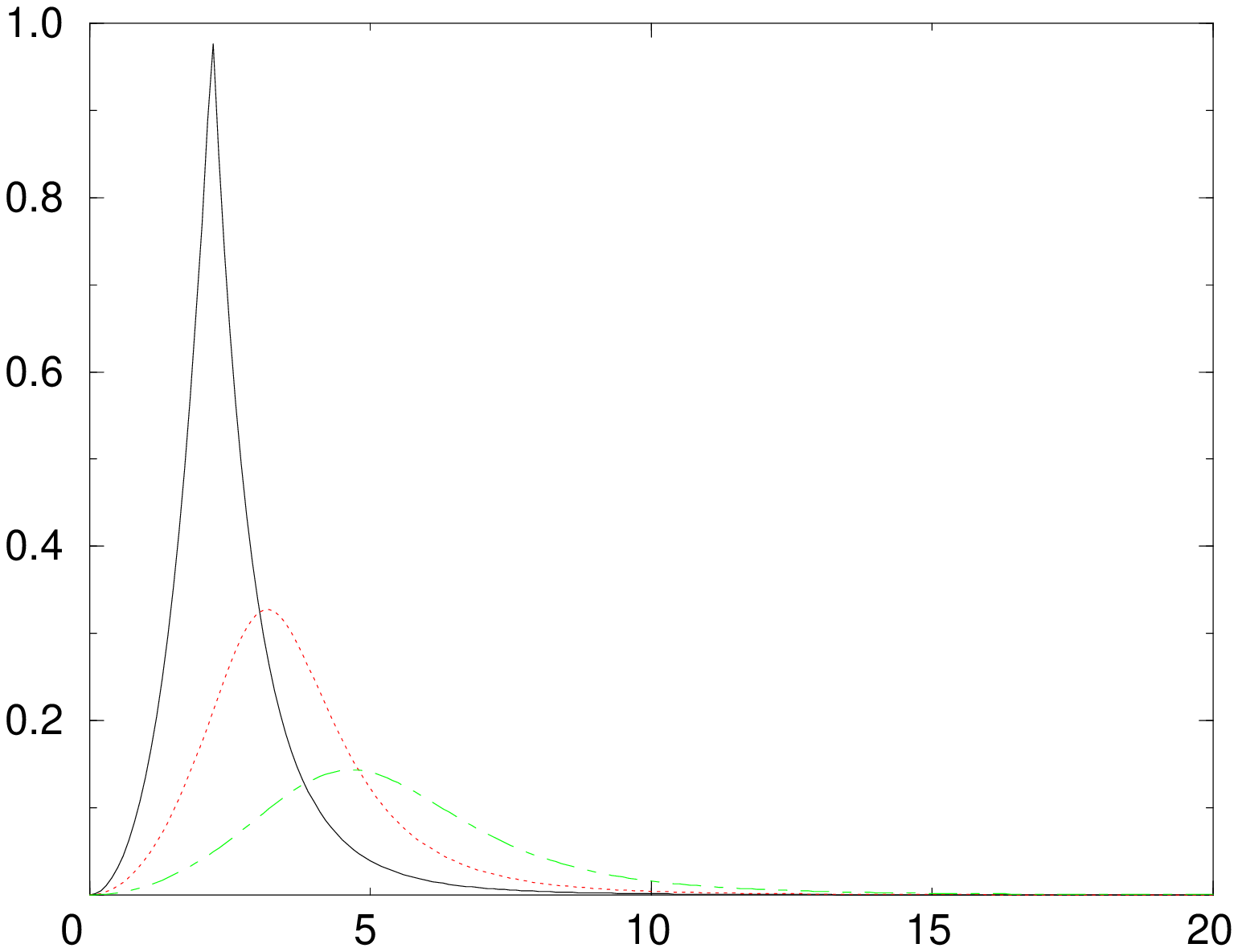,width=10cm,height=08cm}
\put(-5,4.8){$ \-- \,{ b_1 \simeq b_{1max}=0.3465} $}
\put(-5,4.3){$ ... \, {b_1=0.2 }$}
\put(-5,3.8){$ $-.-$ \,{ b_1=0.1} $}
\put(-12.2,4){$- \frac{2}{hv^2} B(r)$}
\put(-4.8,-.38){$r$}
\end{picture}
\vskip 0.3cm
\caption{The magnetic field $B(r)$ as a function of $r$ for $n=1$ and
several values of $b_1$ in the non-topological phase. The vortices are ring
type as in the topological sector.}
\end{figure}

%%%%%%%%%%%%%%%%%%%%%%% Endfigures-Nontopological%%%%%%%%%%%%%%%%%%%%%%


\begin{thebibliography}{99}

\bibitem{anomalo}  I.I. Kogan, Phys. Lett. {\bf B262} (1991) 83; J. Stern,
Phys. Lett. {\bf B265} (1991) 119.

\bibitem{lee}  B.-H. Lee, C. Lee, and H. Min, Phys. Rev. {\bf D45} (1992)
4588.

\bibitem{spector}  Z. Hlousek and D. Spector, Nucl. Phys. {\bf B370} (1992)
143, {\it ibid.} {\bf B397} (1993) 173.

\bibitem{nos}  H.R. Christiansen, M.S. Cunha, J.A. Helay\"{e}l-Neto, L.R.U.
Manssur, A.L.M.A. Nogueira, {\it $N=2$ Maxwell-Chern-Simons model with
anomalous magnetic moment coupling via dimensional reduction},
Int. J. Mod. Phys. {\bf A14} (1999) 147.

\bibitem{navratil} P. Navr\'{a}til, Phys. Lett. {\bf B365} (1996) 119.

\bibitem{laplata} J.D. Edelstein, C. N\'{u}\~{n}ez and F.A. Schaposnik,
Phys.Lett.{\bf B329} (1994) 39.

\bibitem{llw} C. Lee, K. Lee, E.J. Weinberg, Phys. Lett. {\bf B243} (1990)
105.

\bibitem{torres}  M. Torres, Phys. Rev. {\bf D46} (1992) R2295; T. Lee and
H. Min, Phys. Rev. {\bf D50} (1994) 7738; T. A. Antillon, J. Escalona and M.
Torres, Phys. Rev. {\bf D55} (1997) 6327.

\bibitem{sohnius}  M.F. Sohnius, Phys. Rep. {\bf 128} (1985) 39.

\bibitem{jw}  R. Jackiw and E.J. Weinberg, Phys. Rev. Lett. {\bf 64}
(1990) 2234.

\end{thebibliography}
\end{document}